\documentclass[twocolumn]{aastex62}

\newcommand\CII{C\,{\scriptsize II}}
\newcommand\CIV{C\,{\scriptsize IV}}
\newcommand\MgII{Mg\,{\scriptsize II}}
\newcommand\Omegab{\Omega_{\rm b}}
\newcommand\OmegaM{\Omega_{\rm M}}
\newcommand\FWHMCII{{\rm FWHM_{CII}}}
\newcommand\FWHMCO{{\rm FWHM_{CO}}}
\newcommand\amin{a_{\rm min}}
\newcommand\amaj{a_{\rm maj}}
\newcommand\MBH{M_{\rm BH}}
\newcommand\MBHmin{M_{\rm BH}^{\rm min}}
\newcommand\MBHmax{M_{\rm BH}^{\rm max}}
\newcommand\MDH{M_{\rm h}}
\newcommand\MDHmin{M_{\rm h}^{\rm min}}
\newcommand\MDHmax{M_{\rm h}^{\rm max}}
\newcommand\Mdyn{M_{\rm dyn}}
\newcommand\Mstar{M_\star}
\newcommand\Mbaryon{M_{\rm b}}
\newcommand\vrot{V_{\rm rot}}
\newcommand\vcirc{V_{\rm circ}}
\newcommand\fbaryon{f_{\rm b}}
\newcommand\Lbol{L_{\rm bol}}
\newcommand\LIR{L_{\rm IR}}
\newcommand\LFIR{L_{\rm FIR}}
\newcommand\BHAR{BHAR}
\newcommand\rvir{r_{\rm vir}}

\graphicspath{{./}{figures/}}

\received{January 15, 2019}
\accepted{February 7, 2019}
\submitjournal{ApJL}

\shorttitle{Rapid Growth of SMBHs}
\shortauthors{Shimasaku and Izumi}

\begin{document}

\title{Black versus Dark: Rapid Growth of Supermassive Black Holes in Dark Matter Halos at $z\sim6$}

\correspondingauthor{Kazuhiro Shimasaku}
\email{shimasaku@astron.s.u-tokyo.ac.jp}

\author[0000-0002-2597-2231]{Kazuhiro Shimasaku}
\affil{Department of Astronomy, School of Science, The University of Tokyo, 
7-3-1 Hongo, Bunkyo, Tokyo 113-0033, Japan}
\affil{Research Center for the Early Universe, School of Science, The University of Tokyo, 
7-3-1 Hongo, Bunkyo, Tokyo 113-0033, Japan}

\author[0000-0001-9452-0813]{Takuma Izumi}
\affiliation{National Astronomical Observatory of Japan, 
2-21-1 Osawa, Mitaka, Tokyo 181-8588, Japan}
\affil{NAOJ fellow}

\begin{abstract}

We report on the relation between the mass of supermassive black holes (SMBHs; $\MBH$) and that of hosting dark matter halos ($\MDH$) for 49 $z\sim6$ quasi-stellar objects (QSOs) with [\CII]158$\mu$m velocity-width measurements. Here, we estimate $\MDH$ assuming that the rotation velocity from $\FWHMCII$ is equal to the circular velocity of the halo; we have tested this procedure using $z\sim 3$ QSOs that also have clustering-based $\MDH$ estimates. 
We find that a vast majority of the $z\sim6$ SMBHs are more massive than expected from the local $\MBH$--$\MDH$ relation, with one-third of the sample by factors $\gtrsim 10^2$. The median mass ratio of the sample, $\MBH/\MDH = 6\times 10^{-4}$, means that $0.4\%$ of the baryons in halos are locked up in SMBHs. 
The mass growth rates of our SMBHs amount to $\sim 10\%$ of the star formation rates (SFRs), or $\sim 1\%$ of the mean baryon accretion rates, of the hosting galaxies. A large fraction of the hosting galaxies are consistent with average galaxies in terms of SFR and perhaps of stellar mass and size. 
Our study indicates that the growth of SMBHs ($\MBH \sim 10^{8-10} M_\odot$) in luminous $z\sim6$ QSOs greatly precedes that of hosting halos owing to efficient gas accretion even under normal star formation activities, although we cannot rule out the possibility that undetected SMBHs have local $\MBH/\MDH$ ratios. This preceding growth is in contrast to much milder evolution of the stellar-to-halo mass ratio.

\end{abstract}



\section{Introduction} \label{sec:intro}

Observations have identified more than 200 supermassive black holes (SMBHs) shining as QSOs in the early universe before the end of cosmic reionization, or $z\gtrsim6$, with the most distant one being located at $z=7.54$ (\citet{Venemans17}) and the most massive ones having order $\sim 10^{10} M_\odot$. How these SMBHs grow so massive in such early epochs remains a topic of debate. To resolve this, it is key to reveal what galaxies host these SMBHs, because SMBHs and galaxies are thought to co-evolve by affecting each other, as is inferred from various correlations between them seen locally (e.g., \citet{Kormendy13} for a review).

At high redshifts like $z\sim6$, the parameters of hosting galaxies that are often examined are central velocity dispersion ($\sigma$) and dynamical mass ($\Mdyn$), with the latter being a proxy of stellar mass ($\Mstar$). The relations between these parameters and black hole mass ($\MBH$) are then compared with the corresponding local relations for ellipticals and bulges. It has been found that the $\MBH$--$\sigma$ relation at $z\sim6$ is not significantly different from the local one (e.g., \citet{Willott17}). On the other hand, $z\sim6$ SMBHs appear to be overmassive compared with local counterparts with the same bulge mass (e.g., \citet{Decarli18}), although faint QSOs are on the local relation (\citet{Izumi18}). Note that these comparisons are not so straightforward because the stellar components of QSOs may not be bulge-like and may also be greatly contaminated by cold gas (e.g., \citet{Venemans17}, \citet{Feruglio18}).

The relation between $\MBH$ and the mass of hosting dark halos ($\MDH$;  \citet{Ferrarese02}) provides different insights into co-evolution, by directly constraining the SMBH growth efficiency in halos. For example, let us assume two cases: (1) that stellar components and SMBHs grow at similarly high paces, or (2) that they grow at similarly low paces. Both cases give similar $\MBH$--$\Mstar$ relations, but the former predicts a higher $\MBH$--$\MDH$ relation. Cold gas in a halo is used for both star formation and SMBH growth, with shares and consumption rates being controlled by various physical processes. The $\MBH$--$\MDH$ relation at high redshifts may lead to the disentangling of some of these processes.

In this Letter, we derive the $\MBH$--$\MDH$ relation for $z\sim6$ QSOs and compare it with the local relation. We also examine the efficiency of SMBH growth by comparing the growth rate with the star formation rate (SFR) of hosting galaxies and the baryon accretion rate (BAR) of hosting halos. We estimate $\MDH$ from [\CII]158$\mu$m line widths, assuming that lines are broadened by disk rotation and that the rotation velocity is equal to the circular velocity of hosting halos. We show that this procedure appears to be valid, using lower-$z$ QSOs.

In Section \ref{sec:data}, we calculate $\MDH$ for a $z\sim6$ QSO sample compiled from the literature. Results are presented and discussed in Section \ref{sec:results}. Concluding remarks are given in Section \ref{sec:conclusions}. We adopt a flat cosmology with $(\OmegaM, \Omegab, \Omega_\Lambda, H_0) = (0.3, 0.05, 0.7, 70\ {\rm km\ s^{-1}\ Mpc^{-1}})$ and the AB magnitude system.


\section{Sample and halo mass estimation} \label{sec:data}

\begin{figure}[h]
\includegraphics[width=0.50\textwidth]{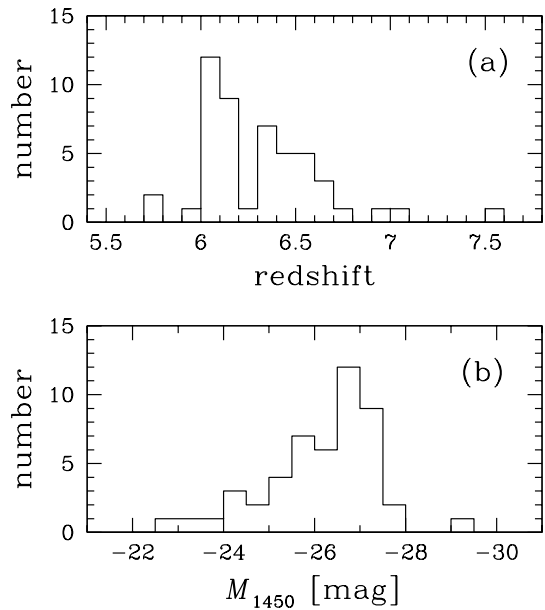}
\caption{Redshift (panel (a)) and $M_{1450}$ (panel (b)) distributions of the sample.
\label{fig:hist_z_M1450}}
\end{figure}

We use 49 $z\sim 6$ QSOs with $\MBH$ and $\FWHMCII$ data
in the literature, where most of the $\FWHMCII$ data were taken with the Atacama Large Millimeter/submillimeter Array (ALMA) at a high spatial resolution.
Among them, 20 have an $\MBH$ measurement based on a broad emission line (\MgII$\lambda 2799$ in most cases), while the remaining 29 have only a minimum $\MBH$ value calculated from the 1450\AA\ luminosity ($L_{1450}$) on the assumption of Eddington-limited accretion
\footnote{Objects with \MgII\ (or \CIV)-based $\MBH$ ($N=20$): J0055$+$0146 (\citet{Willott15}), J0100$+$2802 (\citet{Wang16}), J0109$-$3047 (\citet{Venemans16}), J0210$-$0456 (\citet{Willott13}), PSOJ036$+$03 (\citet{Banados15}), J0305$-$3150 (\citet{Venemans16}), J1044$-$0125 (\citet{Wang13}), J1120$+$0641 (\citet{Venemans12}), J1148$+$5251 (\citet{Walter09}), J1342$+$0928 (\citet{Venemans17}), PSOJ323$+$12 (\citet{Mazzucchelli17}), J2100$-$1715, J2229$+$1457 (\citet{Willott15}), J0338$+$29 (\citet{Mazzucchelli17}), J2329$-$0301 (\citet{Willott17}), J2348$-$3054 (\citet{Venemans16}), PSOJ167$-$13 (\citet{Venemans15}), PSOJ231$-$20 (\citet{Mazzucchelli17}), J0859$+$0022 (\citet{Izumi18}), J2216$-$0010 (\citet{Izumi18})).
Those without ($N=29$): J0129$-$0035 (\citet{Wang13}), J1319$+$0950 (\citet{Wang13}), J2054$-$0005 (\citet{Wang13}), VMOS2911 (\citet{Willott17}), J2310$+$1855 (\citet{Wang13}), J1152$+$0055 (\citet{Izumi18}), J1202$-$0057 (\citet{Izumi18}), and 22 objects given in Table 2 of \citet{Decarli18} after excluding those without $\FWHMCII$ data and PSOJ231$-$20.}. The systematic uncertainty in broad line-based $\MBH$ estimates is $\sim 0.5$ dex (e.g., \citet{Shen13} for a review of $\MBH$ estimation).
Shown in Figure \ref{fig:hist_z_M1450} are the redshift and rest frame $1450$\AA\ absolute magnitude ($M_{1450}$) distributions of the 49 objects.

For each object, we calculate the rotation velocity as $\vrot = 0.75 \FWHMCII / \sin i$ following \citet{Wang13}, assuming that the [\CII] line is broadened solely by disk rotation. Here, $i = \cos^{-1} \amin/\amaj$ is the inclination angle of the disk, with $\amin$ and $\amaj$ being the minor and major axes, respectively, of the deconvolved [\CII] image. We set $i=55^\circ$ (average value for randomly inclined disks) when $\amin/\amaj$ data are unavailable (e.g., \citet{Willott17})\footnote{The average value of the objects with $\amin/\amaj$ data is $52^\circ$.}. 
We then assume that $\vrot$ is equal to the circular velocity of the hosting dark matter halo, $\vcirc$, and convert $\vcirc$ into $\MDH$ using the spherical collapse model (equation [25] of \citet{Barkana01}).

This procedure to derive $\MDH$ from $\FWHMCII$ contains several assumptions that cannot be completely verified by current data.
One is that [\CII] emitting regions are rotating disks. A velocity gradient has been found for several QSOs (e.g., \citet{Wang13},\citet{Willott13}). With high-resolution ALMA data, \citet{Shao17} have derived a rotation curve of the $z=6.13$ QSO ULAS J1319$+$0950 that is flat at $\gtrsim 1.5$ kpc radii. This object is included in our sample, and we find that the calculated $\vrot$ agrees with the flat rotation velocity. On the other hand, \citet{Venemans16} have ruled out a flat rotation for QSO J0305$-$3150. In any case, the number of QSOs with high-quality [\CII] data is still very limited.
We note that if we assume that [\CII] line widths are solely due to random motion and if velocity dispersion $\sigma (= {\rm FWHM}/2.35)$ is converted into $\vcirc$ by $\vcirc = \sqrt{2} \sigma$, we obtain lower $\vcirc$ and hence lower $\MDH$ because of $\sqrt{2}/2.35 < 0.75$.
As found in Section \ref{sec:results}, adopting lower $\MDH$ values enlarges the offset of our QSOs from the local $\MBH$--$\MDH$ relation.

Another key assumption that cannot be tested is $\vrot = \vcirc$. While local spiral galaxies have $\vrot/\vcirc \simeq 1.2$--$1.4$, it is not clear whether high-$z$ QSO host galaxies have also similarly high ratios; if they have such high ratios, our procedure will be overestimating $\MDH$ by a factor of $1.2^3$--$1.4^3\simeq 2$--$3$. On the other hand, \citet{Chen18} have shown $\vrot/\vcirc > 0.6$ by imposing that the duty cycle defined as the ratio of the number density of $z\sim6$ QSOs to that of hosting dark halos has to be less than unity. 

We cannot thoroughly verify the assumptions one by one, so we indirectly test our procedure as a whole by comparing $\MDH$ derived from our procedure with those based on clustering analysis at high redshifts. We do so at $z<6$ as there is no clustering study at $z\gtrsim6$. The best sample for this test is \citet{Trainor12}'s $z=2.7$ sample, for which both a clustering-based $\MDH$ estimate and FWHM data are available.
\citet{Trainor12} have obtained a median halo mass of 15 QSOs at $z=2.7$ to be $\MDH = 10^{12.3 \pm 0.5} M_\odot$ from cross-correlation with galaxies around them. Among them, 12 have CO(3$\rightarrow$2) velocity-width measurements by \citet{Hill19}
\footnote{Since CO and [\CII] lines trace different regions of a galaxy, we check if they give similar FWHM values, using eight objects from our sample with CO(6$\rightarrow$5) FWHM measurements. We find that $\FWHMCII$ is $7\%$ smaller than $\FWHMCO$ on average, but this difference is not statistically significant when the errors in both measurements are considered. (The mean relative errors in $\FWHMCII$ and $\FWHMCO$ are $11\%$ and $25\%$, respectively.)}.
We apply our procedure to nine of the 12 objects (after excluding three with a complex line profile), finding $\MDH = 10^{12.14}$--$10^{13.17}M_\odot$ with a median of $10^{12.71} M_\odot$. This median value is consistent with that from the clustering analysis within the $1\sigma$ error in the latter. See Table \ref{tab:mdh_comp} for a summary of the comparison.

As an additional but less stringent test, we compare $\MDH$ of QSOs at $z\sim 4.5$ with clustering results at similar redshifts. Here, $z\sim4.5$ is the lowest redshift at which the \CII\ line is accessible from the ground, and roughly corresponds to the maximum redshift where clustering data are available.
We use nine QSOs with $\FWHMCII$ data (\citet{Wagg10},\citet{Wagg12}, \citet{Trakhtenbrot17}), and find that their masses are in the range $10^{11.46} < \MDH/M_\odot < 10^{13.44}$ with the median $10^{12.34} M_\odot$. This mass range is comparable to $\MDH$ estimates for a large number of $z\sim 3$--$5$ QSOs from correlation analysis, $10^{12.15}$--$10^{13.18} M_\odot$ (\citet{Shen09}, \citet{He18}, \citet{Timlin18}; Table \ref{tab:mdh_comp}). We regard this rough agreement as modest support for our procedure, because the $\MDH$ range of the $z\sim4.5$ QSOs is broad and because FWHM-based and clustering-based masses are compared for different samples.

\begin{table}
    \caption{Comparisons between clustering-based and FWHM-based $\MDH$.}
    \begin{tabular}{cccc}
    \hline
     $z$ & $\log \MBH$ [$M_\odot$] & \multicolumn{2}{c}{$\log \MDH$ [$M_\odot$]} \\
     \cline{3-4}
               &        & Clustering & FWHM \\
    \hline
     $2.7$     & $8.8$--$9.7^{a)}$ 
               & \underline{$12.3\pm0.5$}$^{a)}$
               & \underline{$12.71$} [$12.14$--$13.17$]\\
     $3$--$5$  & $7.8$--$10.0^{b)}$
               & $12.15$--$13.18^{b)}$
               & ... \\
     $4.5$     & $8.4$--$9.8^{c)}$ 
               & ...
               & \underline{$12.34$} [$11.46$--$13.44$] \\
    \hline
    \end{tabular}
    
    Notes. Underscored numbers mean the median value, while others correspond to the full range over the sample. The $\MBH$ of the $z\sim 4.5$ sample are based on broad emission lines, while those of the other samples are calculated from $L_{1450}$ on the assumption of the Eddington-limited accretion. References. $(a)$  \citet{Trainor12}, $(b)$ \citet{He18}, \citet{Shen09}, \citet{Timlin18}, $(c)$ \citet{Wagg10}, \citet{Wagg12}, \citet{Trakhtenbrot17}.
    \label{tab:mdh_comp}
\end{table}

These comparisons indicate that this procedure can be used as a rough estimator of $\MDH$ at least in the statistical sense, although the evaluation of its uncertainty is limited by that in \citet{Trainor12}'s mass estimate. Our procedure gives a 0.4 dex higher median mass than that of \citet{Trainor12}. However, because this difference is within the $1\sigma$ error in their estimate, $0.5$ dex, we do not correct our procedure for this possible systematic overestimation. The comparison also indicates that the underestimation by this procedure, if any, appears modest, $< 0.5$ dex.
Our main result that the SMBHs in $z\sim6$ QSOs have higher $\MBH/\MDH$ than local values is robust, because this result holds as long as the systematic underestimation of $\MDH$ is $\lesssim 0.5$ dex.

The $\MDH$ values of our $z\sim6$ QSOs thus obtained are less than $1\times 10^{13} M_\odot$ except for two objects. The median of the entire sample is $1.2\times 10^{12} M_\odot$, with a central $68\%$ range of $(0.6$--$3.4) \times 10^{12} M_\odot$. These relatively low masses are consistent with the halo mass distribution of $z\sim6$ QSOs constrained from the statistics of companion galaxies by \citet{Willott05}.


\section{Results and Discussion} \label{sec:results}

\subsection{Mass vs. mass}

\begin{figure}
\includegraphics[width=0.50\textwidth]{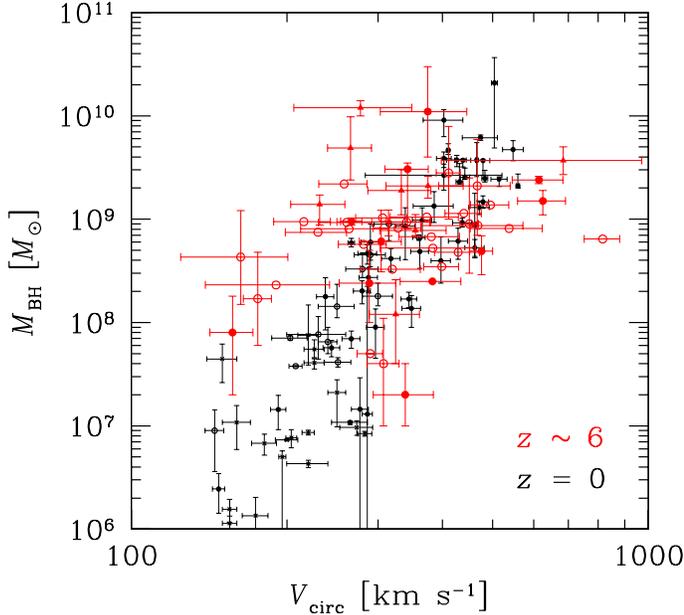}
\caption{$\MBH$ plotted against $\vcirc$. The red symbols indicate $z\sim 6$ QSOs. Filled symbols mean broad line-based $\MBH$ measurements, while open symbols indicate minimum values on the assumption of Eddington accretion. Circles are objects with an inclination angle measurement, while triangles are those without; for the latter, $i=55^\circ$ is assumed. Black symbols are local galaxies taken from \citet{Kormendy13}: filled circles, ellipticals; open circles, classical bulges; crosses, pseudo bulges.
\label{fig:mbh_vcirc}}
\end{figure}

\begin{figure*}
 \gridline{\fig{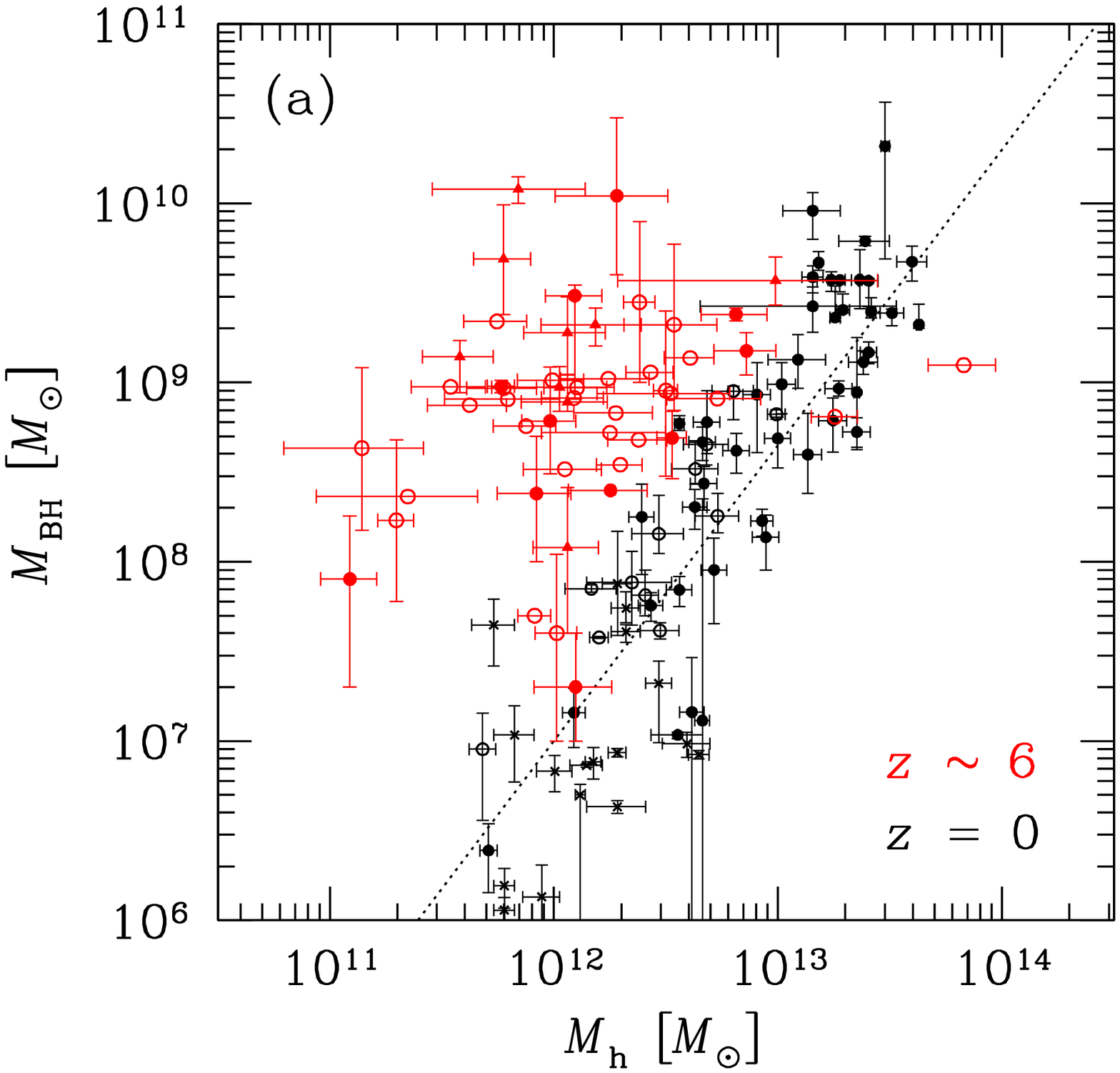}{0.50\textwidth}{}
           \fig{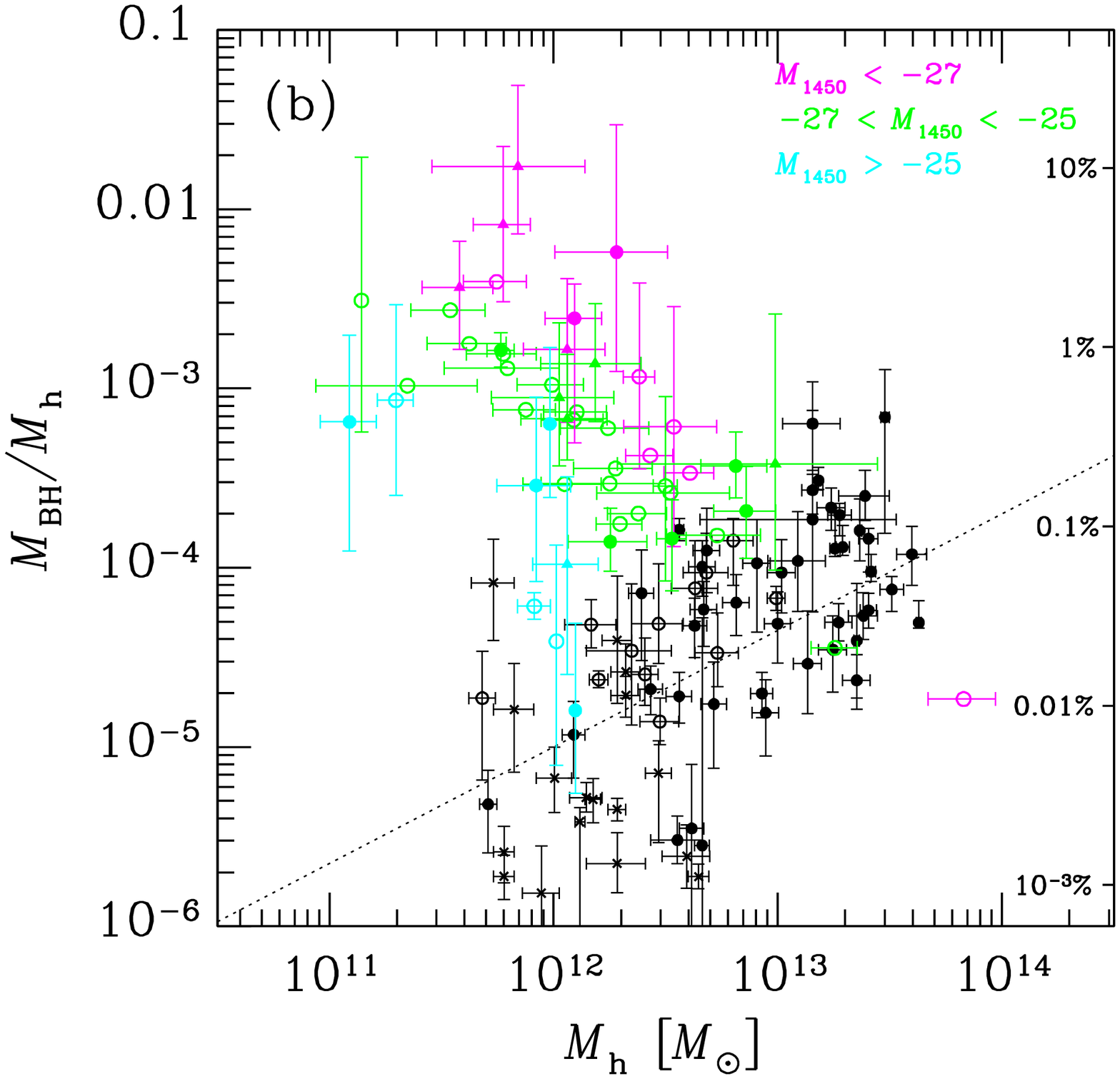}{0.50\textwidth}{}
           }
 \caption{$\MBH$ (panel [a]) and $\MBH/\MDH$ ([b]), plotted against $\MDH$. The meanings of the symbols are the same as in Figure \ref{fig:mbh_vcirc}. Dotted lines are the local relation obtained by \citet{Ferrarese02}. In panel (b), $z\sim6$ objects are colored depending on $M_{1450}$: magenta, brighter than $-27$; green, $-27$ to $-25$; cyan, fainter than $-25$. The $y$ axis of the right-hand side of panel (b) indicates the fraction of baryons in halos that are locked up in SMBHs.
 \label{fig:mbh_ratio_mdh}}
 \end{figure*}

Figure \ref{fig:mbh_vcirc} shows $\MBH$ against $\vcirc$ for the 49 $z\sim 6$ QSOs, together with local galaxies taken from \citet{Kormendy13} for which we convert central velocity dispersions into $\vcirc$ using the formula given in \citet{Pizzella05}. The very weak correlation seen in the $z\sim6$ sample is partly due to large intrinsic errors in both $\MBH$ and $\vcirc$. If the observed values are taken at face value, about two-thirds of the $z\sim 6$ QSOs are consistent with the distribution of local galaxies, while the remaining one-third have higher $\MBH$. 

Figure \ref{fig:mbh_ratio_mdh} plots $\MBH$ versus $\MDH$. In contrast to Figure \ref{fig:mbh_vcirc}, most of the $z\sim6$ QSOs deviate from the local relation (\citet{Ferrarese02}) toward higher $\MBH$, or lower $\MDH$. This is because $\MDH$ at a fixed $\vcirc$ decreases with redshift as $(1+z)^{-1}$.
Most of the $z\sim 6$ QSOs have a $\gtrsim 10$ times more massive SMBH than local counterparts with the same $\MDH$, with one-third by factor $\gtrsim 10^2$. Thus, at $z\sim6$ the growth of SMBHs precedes that of hosting halos at least for most luminous QSOs. This is in contrast to a roughly redshift-independent $\Mstar$--$\MDH$ relation of average galaxies (e.g., \citet{Behroozi18}).

The overmassive trend observed here may be due to selection effects because the sample is biased for luminous QSOs (e.g., \citet{Schulze14}). We cannot rule out the possibility that SMBHs at $z\sim6$ are in fact distributed around the local relation with a large scatter and that we are just observing its upper envelope truncated at $\MDH\sim 10^{13} M_\odot$ beyond which objects are too rare to find because of an exponentially declining halo mass function (for the halo mass function, see, e.g., \citet{Murray13}). The results obtained in this study apply only to luminous QSOs detectable with current surveys.

The median $\MBH/\MDH$ ratio of the entire sample is $6.3\times 10^{-4}$ with a central $68\%$ tile of $1.5\times 10^{-4}$--$1.8\times 10^{-3}$. Even when limited to the objects with relatively reliable $\MBH$ and $\MDH$ data shown by red filled circles, we find a large scatter in $\MBH$ at a fixed $\MDH$, suggesting a wide spread in SMBH growth efficiency. 
We calculate the fraction of baryons in the hosting dark halo that are locked up in the SMBH, as $\fbaryon = \MBH /\Mbaryon$, where $\Mbaryon \equiv (\Omegab/\OmegaM)\MDH$ is the total mass of baryons in a halo. Our sample has a median $\fbaryon$ of $0.4 \%$, with some well above $1\%$.

In Figure \ref{fig:mbh_ratio_mdh}(b), QSOs with brighter $M_{1450}$ magnitudes tend to have higher $\MBH/\MDH$ ratios. This trend appears to be reasonable because at a given $\MDH$, those with a higher $\MBH$ can be brighter because the Eddington luminosity is proportional to $\MBH$.
Note that some of the faint objects also have very high ratios, far above the local values.

\begin{figure}[ht!]
\includegraphics[width=0.50\textwidth]{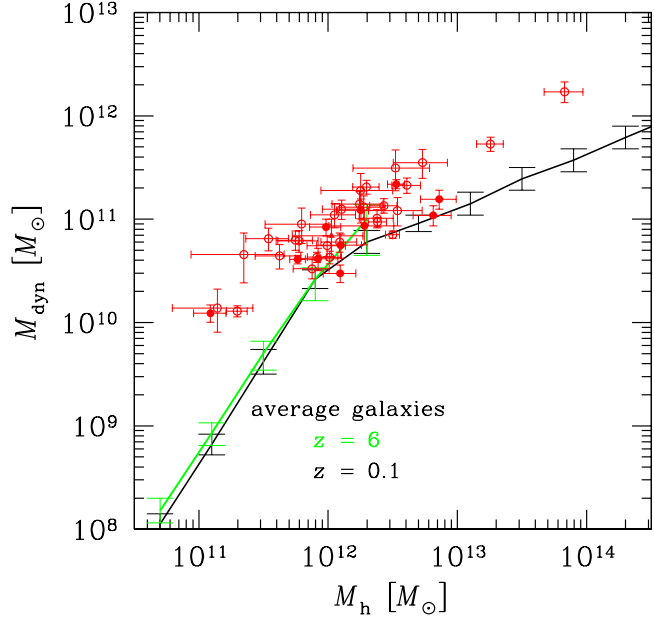}
\caption{$\Mdyn$ plotted against $\MDH$. The meanings of the symbols are the same as in Figure \ref{fig:mbh_vcirc}. Lines with errors indicate the relations for average galaxies at $z=6$ (green) and $z=0.1$ (black) given in \citet{Behroozi18}; the $z=6$ relation at $\MDH>2\times 10^{12} M_\odot$ has not been constrained.
\label{fig:mdyn_mdh}}
\end{figure}

We compare $\Mdyn$ with $\MDH$ for 41 objects with size data in Figure \ref{fig:mdyn_mdh}\footnote{We use $\Mdyn/M_\odot = 1.16\times 10^5 (\vrot/{\rm km\ s^{-1}})^2 (D/{\rm kpc})$, with $D=1.5\amaj$ (\citet{Willott15}). In this definition of $\Mdyn$, $\Mdyn$ vs. $\MDH$ is essentially equivalent to $D$ vs. $\vcirc$ if $\vrot=\vcirc$.}, finding a nearly linear correlation with a median ratio of $\Mdyn/\MDH = 0.07$ (central $68\%$: $0.04-0.10$). 
Although our objects are distributed nearly a factor of two above the relation of $z=6$ average galaxies \citep{Behroozi18}, the difference is probably insignificant when various uncertainties in these quantities are considered. For example, $\Mdyn$ may be significantly contaminated by molecular gas mass as reported for some QSOs (e.g., \citet{Venemans17}, \citet{Feruglio18}).

We also compare the [\CII] emission radii of our objects with the virial radii ($\rvir$) of the hosting halos ($\rvir=G\MDH/\vcirc^2$ where $G$ is the gravitational constant), finding a median ratio of 0.04 (central $68\%$: $0.02-0.07$). This result appears to be consistent with rest-ultraviolet (UV) effective radius-to-$\rvir$ ratios, typically $\sim 0.03$, obtained for $z\sim6$ galaxies (\citet{Kawamata18}), suggesting that galaxies hosting $z\sim6$ QSOs do not have extreme sizes.

\begin{figure}[ht!]
\includegraphics[width=0.50\textwidth]{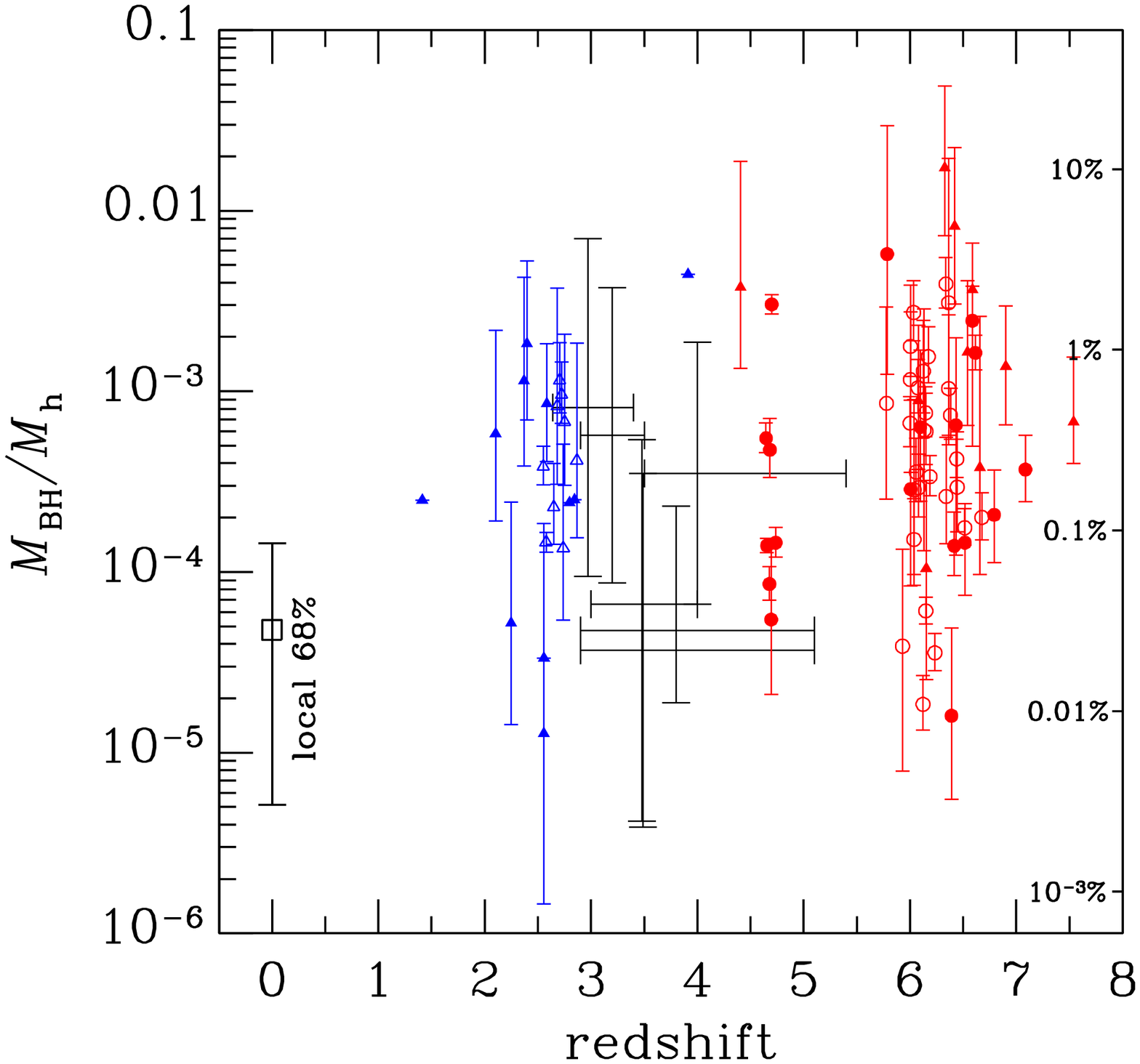}
\caption{$\MBH/\MDH$ against redshift. The $\MDH$ of colored objects are derived by our procedure from $\FWHMCII$ (red) and $\FWHMCO$ (blue: \citet{Coppin08}, \citet{Hill19}, \citet{Shields06}). Red symbols at $z\sim 4.5$ are the objects used to test our procedure in Section \ref{sec:data}. Black error bars are constraints from clustering analysis; for each data point, the vertical errors correspond to the range $\MBHmin/\MDHmax < \MBH/\MDH < \MBHmax/\MDHmin$ where $\MDHmin$ ($\MDHmax$) is the $1\sigma$ lower (upper) limit of $\MDH$ inferred from clustering analysis for the given QSO sample, while $\MBHmin$ ($\MBHmax$) is the minimum $\MBH$ derived from the faintest (brightest) $L_{1450}$ of the sample; the horizontal errors correspond to the redshift range of the sample. An open square with errors at $z=0$ indicates the median and the central $68\%$ tile for the local galaxies.
\label{fig:ratio_z}}
\end{figure}

Figure \ref{fig:ratio_z} shows $\MBH/\MDH$ as a function of $z$ for our sample and several supplementary QSO samples at lower redshifts (whose UV magnitudes are distributed in the range $-23.0 > M_{1450} > -29.5$). This figure indicates that luminous QSOs at $z>2$ tend to have overmassive SMBHs irrespective of redshift. We also see a rough agreement of $\MBH/\MDH$ between the clustering-based and FWHM-based results.
Note that the lower-$z$ QSOs plotted here are unlikely to be descendants of the $z\sim6$ QSOs because QSOs' lifetimes, typically $\sim 10^{6-8}$ yr (e.g., \citet{Martini04}), are much shorter than the time intervals between $z\sim6$ and these lower redshifts.

\subsection{Growth rate vs. growth rate}

We then compare the mass growth rate of SMBHs with the SFR and the mean BAR of hosting halos ($\langle$BAR$\rangle$); we use $\langle$BAR$ \rangle$ because halos at a fixed $\MDH$ can take a wide range of BAR values (e.g., \citet{Fakhouri10}) and we cannot tell what value each of our objects actually has. For this comparison, we only use 18 objects with broad line-based $\MBH$ data and infrared (IR) luminosity data\footnote{Fifteen objects from \citet{Decarli18}, two from \citet{Izumi18}, and one (J2100$-$1715) from \citet{Walter18}. In the calculation of IR luminosities, a dust temperature of $T_{\rm d}=47$ K and a dust emissivity power-law spectral index of $\beta=1.6$ have been assumed except for J2100$-$1715 for which \citet{Walter18} have obtained $T_{\rm d}=41$ K.}.
SMBH mass growth rates (black hole accretion rates: BHARs) are calculated from $L_{1450}$ as BHAR $= \frac{1-\epsilon}{\epsilon} \Lbol/c^2$, where $\epsilon=0.1$ (fixed) is the mass-energy conversion efficiency, and $\Lbol$ is the bolometric luminosity estimated using the formula:  
$\Lbol/{\rm erg\ s^{-1}} = 10^{4.553} L_{1450}^{0.911}/{\rm erg\ s^{-1}}$ (\citet{Venemans16}).
SFRs are obtained from IR luminosities using \citet{Kennicutt12}'s conversion formula: SFR$/M_\odot{\rm yr}^{-1} = 1.49\times 10^{-10} \LIR/L_\odot$. Mean BARs $\langle$BAR$\rangle =(\Omegab/\OmegaM)\langle d\MDH/dt\rangle$ are calculated using the formula given in \citet{Fakhouri10}. \citet{Fakhouri10} have obtained $\langle d\MDH/dt\rangle$ at a given $\MDH$ and a given $z$ from the mean growth of $\MDH$ over a small time step calculated from main branches of merger trees constructed from the Millennium and Millennium II $N$-body simulations.

\begin{figure*}
\gridline{\fig{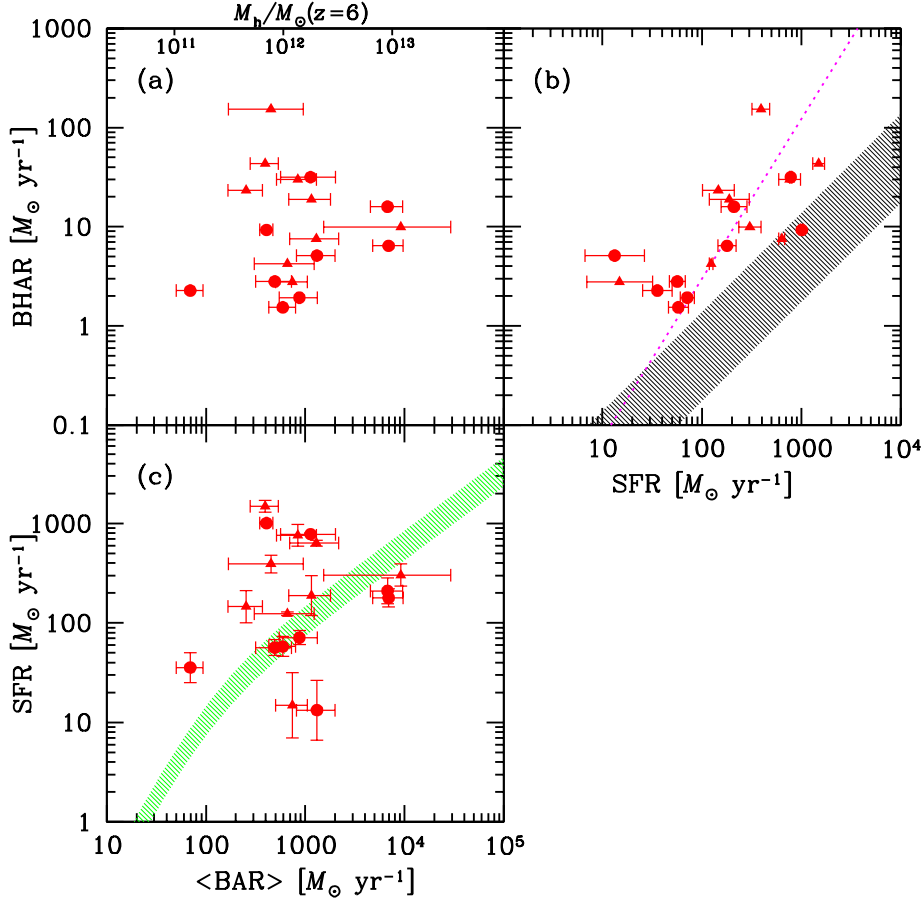}{0.80\textwidth}{}
          }
\caption{Relations between BHAR, SFR, and $\langle$BAR$\rangle$. Panel (a): BHAR vs. $\langle$BAR$\rangle$. Panel (b): BHAR vs. SFR. Panel (c): SFR vs $\langle$BAR$\rangle$. The meanings of the symbols are the same as in Figure \ref{fig:mbh_ratio_mdh}. The upper horizontal axis of panel (a) indicates $\MDH$ corresponding to $\langle$ BAR$\rangle$ at $z=6$. In panel (b), a magenta dotted line indicates the $\LFIR$--$\Lbol$ relation of stacked QSOs at $2<z<7$ by \citet{Wang11}, and a shaded region shows the range of the black hole-to-bulge mass ratio (including an 0.28 dex intrinsic scatter) of local galaxies with bulge masses $10^{10-12} M_\odot$ \citep{Kormendy13}. The $\MBH/\Mstar$ range of local galaxies is not plotted because they are distributed in a much wider range, $-4.0 \lesssim \log \MBH/\Mstar \lesssim -2.0$, depending on $\Mstar$ and morphology. A green shaded region in panel (c) shows the average relation (with $\pm0.15$ dex scatter) of $z\sim6$ galaxies by \citet{Harikane18}.
\label{fig:ar3}}
\end{figure*}

Figure \ref{fig:ar3}(a) plots BHAR against $\langle$BAR$\rangle$. With a large scatter, our QSOs have high BHAR$/\langle$BAR$\rangle$ ratios with a median of $0.6\%$. 
\citet{Yang18} present time-averaged BHARs as a function of $\MDH$ over $0.5<z<4$ using the X-ray luminosity function down to $L_X = 10^{43}$ erg s$^{-1}$ combined with the stellar mass function and the $\Mstar$--$\MDH$ relation. 
Their study covers $44 < \log \Lbol\ {\rm [erg\ s^{-1}]} \lesssim 48.5$, including 2 dex fainter objects than our sample, which is in the range $46.0 < \log \Lbol\ {\rm erg\ s^{-1}} < 48.0$. In their BHAR calculation, all galaxies at given $\Mstar$ are considered. Their results give much lower  BHAR$/\langle$BAR$\rangle \sim 2 \times 10^{-5}$--$1 \times 10^{-4}$ for $\MDH=10^{12}$--$10^{13} M_\odot$ roughly independent of redshift. 
If we assume that $z\sim6$ counterparts to their galaxies also have similarly low time-averaged BHAR/$\langle$BAR$\rangle$ values, then it is implied that the SMBHs of our QSOs are growing $\sim 10^2$ times more efficiently than of average galaxies, maybe being in one of many short growth phases as suggested by \citet{Novak11}.

In Figure \ref{fig:ar3}(b), BHAR correlates with SFR relatively well with a typical ratio of BHAR/SFR $\sim 10\%$, although the correlation may be artificial due to selection effects (\citet{Venemans18}). This ratio is close to those from the average relation of bright QSOs at $2<z<7$ by \citet{Wang11} (dotted line), but higher than the $\MBH/\Mstar$ of local galaxies. Hence, such high ratios should last only for a short period of cosmic time.

Figure \ref{fig:ar3}(c) is a plot of SFR versus $\langle$BAR$\rangle$, showing that our QSOs are distributed around the average relation of $z\sim6$ galaxies (e.g., \citet{Behroozi13}, \citet{Harikane18}), or SFR $\approx 0.1 \langle$BAR$\rangle$, but with a very large scatter. About an half of the objects are consistent with average galaxies. Objects far above the average relation may be starbursts due, e.g., to galaxy merging (when BAR also increases temporarily); the BHAR of these objects is as high as $\sim 0.1 \langle$BAR$\rangle$. 

Finally, we compare the specific growth rates of SMBHs, dark halos, and stellar components.
The 18 SMBHs grow at $\sim 0.1$--$1$ times of the Eddington limit accretion rate, with $\BHAR/\MBH$ being comparable to or higher than the specific halo growth rate, $\langle$BAR$\rangle/\Mbaryon$; the SMBHs are growing faster than the hosting halos on average.
We also find the BHAR$/\MBH$ to be comparable to the specific SFR ($=$SFR$/0.1\Mbaryon$) but with a large scatter \footnote{We have assumed that $10\%$ of baryons are in stars.}. This means that for $z\sim6$ QSOs, SMBHs and stellar components grow at a similar pace on average, confirming the result obtained by \citet{Feruglio18} using $\Mdyn$.

\section{Concluding Remarks} \label{sec:conclusions}

We have estimated $\MDH$ for $49$ $z\sim6$ QSOs from $\FWHMCII$. This procedure appears to be valid as a rough estimator.

We have found that the SMBHs of luminous $z\sim6$ QSOs are greatly overmassive with respect to the local $\MBH$--$\MDH$ relation. This is contrasted with a much milder evolution of the $\Mstar$--$\MDH$ relation of average galaxies over $z\lesssim6$. We have also found that our SMBHs are growing at high paces, amounting to $10^{-1}$SFR, or $10^{-2}\langle$BAR$\rangle$, and that the SFR of hosting galaxies is widely scattered around the SFR--$\langle$BAR$\rangle$ relation of average galaxies.
A large fraction of the hosting galaxies appear to be consistent with average galaxies in terms of SFR, stellar mass, and size, although this result is relatively sensitive to the accuracy of $\MDH$ estimates.

Our study indicates that at $z\sim6$ the growth of SMBHs in luminous QSOs greatly precedes that of hosting halos owing to efficient mass accretion under a wide range of star formation activities including normal star formation, although the existence of faint, undetected SMBHs consistent with the local $\MBH$--$\MDH$ relation cannot be ruled out. These high mass growth paces can last for only a short period, in order to be consistent with the relatively low $\MBH/\MDH$ and $\MBH/\Mstar$ values of local galaxies.

The trend that SMBHs at $z\sim6$ are overmassive vanishes if we are underestimating $\MDH$ by factor 10. Although there is currently no hint of such underestimation, future tests of the procedure using high-$S/N$ [\CII] data and clustering analysis will be useful. 
Simulation studies of the internal structure of high-$z$ galaxies may also be helpful\footnote{\citet{Lupi19} have performed a very high-resolution simulation of a $z=7$ QSO and virtually measured its [\CII] emission by mimicking ALMA observations. Applying our procedure to measured $\FWHMCII$ gives $\MDH = 1.3$--$2.3 \times 10^{12} M_\odot$ depending on the degraded angular resolution, being consistent with the correct value, $1.5 \times 10^{12} M_\odot$. Note also that the host galaxy has a rotating gas disk.}.

SMBH evolution has been implemented in many state-of-the-art galaxy formation models, while detailed comparison with our results is beyond the scope of this Letter. An increasing trend of $\MBH/\MDH$ with redshift is seen in the semi-analytical model by \citet{Shirakata19} (H. Shirakata, private communication). Some hydrodynamical simulations show that $\MDH \sim 10^{12} M_\odot$ halos can have an SMBH as massive as $\sim 10^9 M_\odot$ (e.g., \citet{Costa14}, \citet{Tenneti19}), but based on only several examples. Our results can be used to calibrate the efficiency of SMBH growth in the early cosmic epoch.

\acknowledgments
We thank the referee, Yoshiki Matsuoka, for the insightful comments that greatly improved the manuscript. We also thank Taira Oogi, Hikari Shirakata, and Rieko Momose for useful discussions.




\end{document}